\documentclass[twocolumn,aps,showpacs,floatfix]{revtex4-1}
\usepackage{amssymb,amsmath}
\usepackage{hyperref,graphicx}
\usepackage{dcolumn}
\usepackage{color}
\newcommand{\bs}[1]{{\boldsymbol{#1}}}
\newcommand{\bk}{\bs{k}}
\newcommand{\bq}{\bs{q}}
\newcommand{\br}{\bs{r}}
\newcommand{\bp}{\bs{p}}
\newcommand{\bv}{\bs{v}}
\newcommand{\av}[1]{\overline{#1}}
\newcommand{\rmd}{\mathrm{d}}
\newcommand{\Tr}{\mathrm{Tr}}
\newcommand{\intdd}[1]{\int \frac{\rmd #1}{(2\pi)^d}}
\newcommand{\intdone}[1]{\int \frac{\rmd #1}{2\pi}}
\newcommand{\bra}[1]{\left\langle #1 \right |}
\newcommand{\ket}[1]{\left| #1 \right\rangle}
\newcommand{\tauB}{\tau_\text{B}} 
\newcommand{\lB}{\ell_\text{B}}

\begin{document}

\title{Coherent Backscattering of Ultracold Matter Waves: Momentum Space Signatures}

\author{Nicolas Cherroret$^1$, Tomasz Karpiuk,$^{2,3}$ Cord A.~M\"{u}ller,$^2$ Beno\^{\i}t Gr\'{e}maud,$^{2,4,5}$,  and Christian Miniatura$^{2,4,6}$}                          
\affiliation{
\mbox{$^1$ Physikalisches Institut, Albert-Ludwigs-Universit\"{a}t Freiburg, Hermann-Herder-Str. 3, D-79104 Freiburg, Germany} \\
\mbox{$^2$ Centre for Quantum Technologies, National University of Singapore, 3 Science Drive 2, Singapore 117543, Singapore} \\
\mbox{$^3$ Wydzia{\l} Fizyki, Uniwersytet w Bia{\l}ymstoku, ul. Lipowa 41, 15-424 Bia{\l}ystok, Poland}\\
\mbox{$^4$ Department of Physics, National University of Singapore, 2 Science Drive 3, Singapore 117542, Singapore} \\
\mbox{$^5$ Laboratoire Kastler Brossel, Ecole Normale Sup\'{e}rieure, CNRS, UPMC; 4 Place Jussieu, 75005 Paris, France} \\
\mbox{$^6$ Institut Non Lin\'{e}aire de Nice, UMR 6618, UNS, CNRS; 1361 route des Lucioles, 06560 Valbonne, France} \\}

\begin{abstract}
Using analytical and numerical methods, it is shown that the momentum distribution of a matter wave packet launched in a random potential exhibits a pronounced coherent backscattering (CBS) peak. By analyzing the momentum distribution, key transport times can be directly measured. The CBS peak can be used to prove that transport occurs in the phase-coherent regime, and measuring its time dependence permits monitoring the transition from classical diffusion to Anderson localization.
\end{abstract}

\pacs{03.75.-b, 05.60.Gg, 42.25.Dd, 72.15.Rn}
\maketitle

Disorder has dramatic effects on the quantum transport of matter. Spatial randomness and phase coherence together can completely suppress diffusion, as demonstrated by the paradigmatic phenomenon of Anderson localization \cite{Anderson1978}. During the past decade, there has been a growing body of evidence for three-dimensional (3D) localization in random media with different types of noninteracting waves: light \cite{Storzer06}, microwaves \cite{Chabanov00} and ultrasound \cite{Hu08}. This ubiquitous and yet elusive phenomenon has sparked considerable interest in the field of ultracold atoms \cite{ColdDisorderRevs, Kuhn, Skipetrov08}. Key experimental achievements include 1D Anderson localization in speckle \cite{Billy08} and quasi-periodic \cite{Roati08} potentials, as well as 3D localization in momentum space with the kicked rotor \cite{Chabe}. Recently, 3D Anderson localization of noninteracting ultracold fermions \cite{Kondov11} and bosons \cite{Jendrzejewski11} in a laser speckle field was reported.

To claim Anderson localization, one needs to discriminate interference-induced absence of diffusion from classical trapping or slow diffusion. This requires evidence for phase-coherent transport. Here, the coherent backscattering  (CBS) phenomenon is of key importance because it arises by interference of waves in random media and measures  mesoscopic phase coherence \cite{Montambaux}. With classical waves, CBS appears as an enhancement of the diffuse intensity reflected off a disordered medium around the backscattering direction, and has been observed in numerous experiments involving light \cite{Wolf85, Labeyrie99}, but also acoustic \cite{Fink97} and seismic waves \cite{Larose04}. The interference causing CBS is also responsible for weak localization, by reducing the diffusion coefficent compared to its phase-incoherent, or classical, value \cite{Montambaux,Kuhn}. In electronic systems, weak localization is invaluable for a careful characterization of phase coherence \cite{Niimi2009}. With cold-atomic clouds expanding in random potentials, however, the diffusion constant extracted from real-space data hardly shows clear evidence of localization corrections, because the cloud contains many different momenta that combine to a rather involved spatial profile \cite{Miniatura2009,RobertSV2010}. 

With this Rapid Communication, we propose to study the dynamics of a matter wave that is launched with an initial momentum larger than its momentum spread in the bulk of a 2D or 3D random potential. Combining a numerical and theoretical analysis, we show that the CBS signal can be directly observed in the momentum distribution and studied as a function of time. Ultracold atoms are an invaluable asset as they offer the unique opportunity to visualize the CBS effect on the momentum distribution directly, measured \emph{inside} the disordered medium. With this setup, one avoids the boundary conditions that severely complicate both theory and experiments of wave scattering by random media \cite{Montambaux,Hartung08}. Moreover, the momentum-space analysis gives immediate access to key mesoscopic parameters such as scattering and transport times. Lastly, we demonstrate that the CBS measurement provides precious information on the phase coherence of the matter wave, and finally permits to monitor the transition from diffusion to localization.

\begin{figure}
\resizebox{2.3in}{1.4in} {\includegraphics{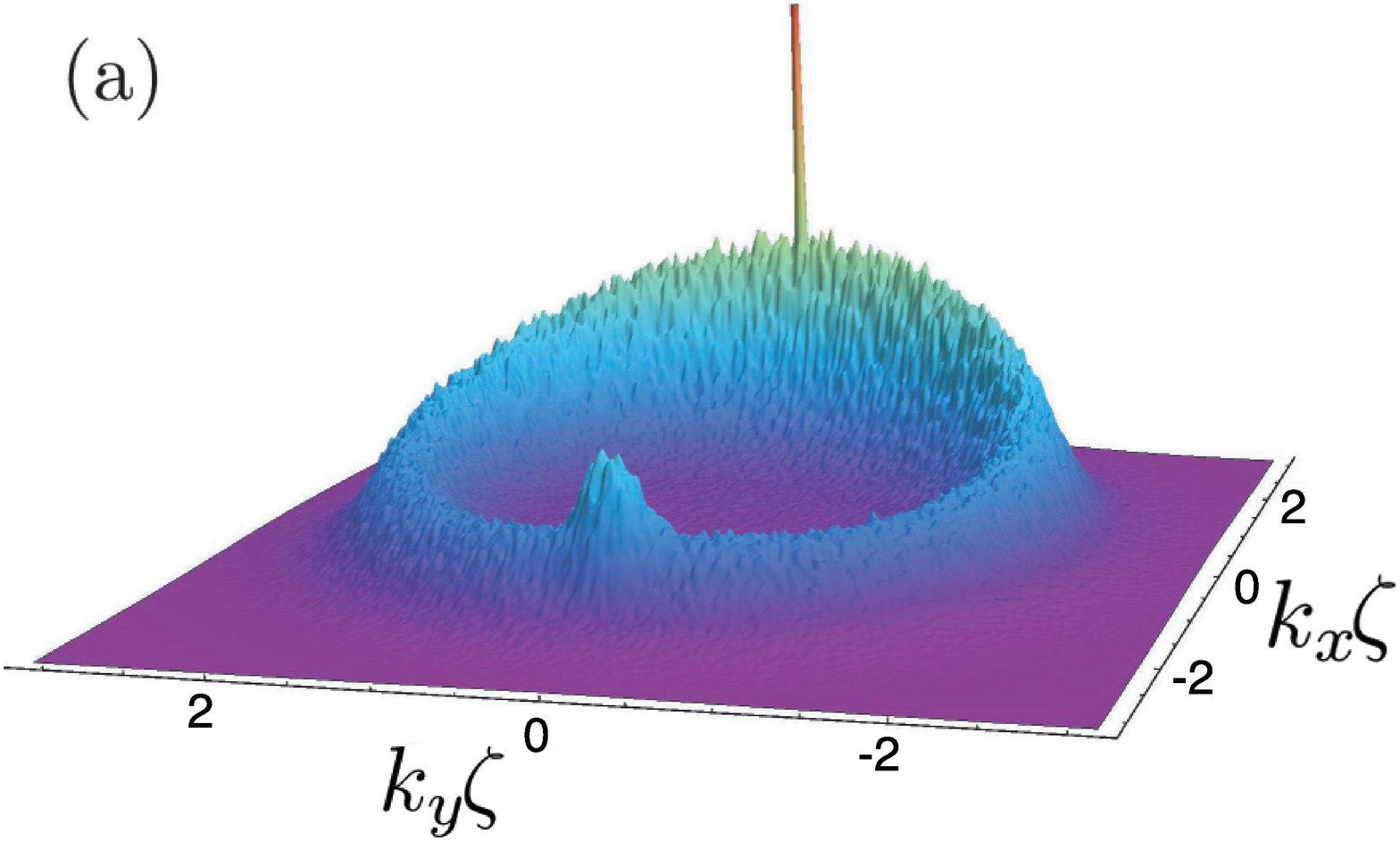}} \\
\resizebox{2.3in}{1.5in} {\includegraphics{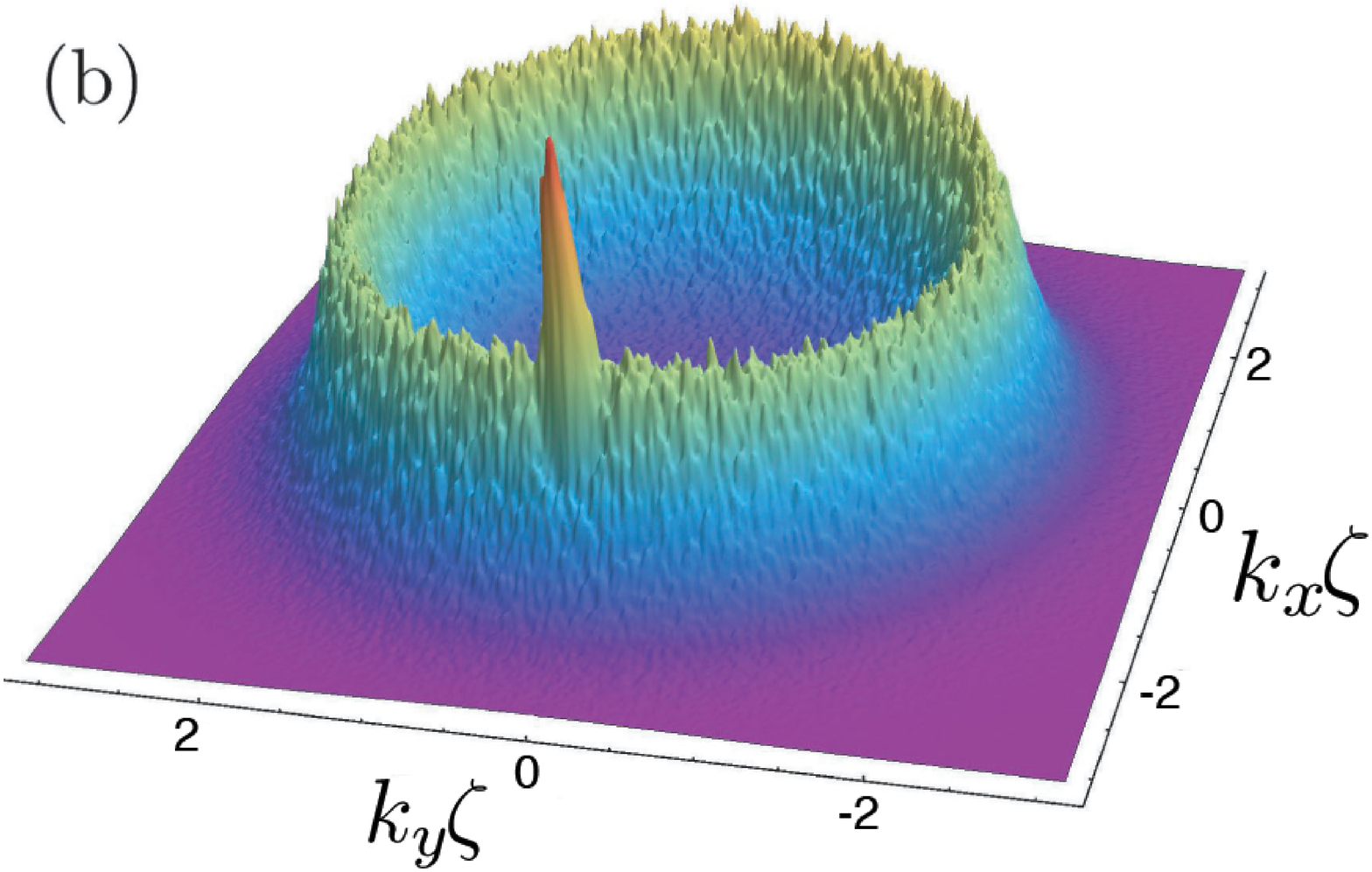}}
\caption{(Color online) Momentum distribution $\overline{\rho}(k_x,k_y,t)$ of a matter wave packet launched with initial momentum $\bk_0 = (k_0,0)$ in a 2D random potential with correlation length $\zeta$, averaged over 960 disorder realizations. The time unit is $\tau_{\zeta}= m\zeta^{2}/\hbar$. (a) $t=10\tau_\zeta$: elastic scattering depletes the initial wave packet, centered at $\bk_0$ (peak values not shown), and populates the disorder-broadened energy shell along the circle $|\bk| = |\bk_0|$, while the CBS peak emerges at $-\bk_0$. (b) $t=18\tau_{\zeta}$: The CBS peak is now the dominant feature, proving phase-coherent multiple scattering.}
\label{cbs}
\end{figure}

Let a cloud of atoms with mass $m$ be prepared at time $t=0$ in the state $\Psi(\bk,0)$ describing a wave packet with mean momentum $\hbar\bk_0=m\bv_0$ and small spread $\Delta k \ll |\bk_0|$. This can be achieved by releasing the atoms from a shallow trap, and either launching them with mean velocity $\bv_0$ or moving the disorder potential with $-\bv_0$ relative to the cloud. We assume negligible interaction effects. This is the case in practice for a very dilute cloud \cite{Billy08} or spin-polarized fermions \cite{Kondov11}. For concreteness, as realized by harmonic trapping of non-interacting particles, we then take the initial distribution $\rho_0(\bk) = |\Psi(\bk,0)|^2$ to be an isotropic Gaussian, 
\begin{equation} 
\label{rho0iso} 
\rho_0(\bk) = (2\pi \Delta k^{-2})^{d/2}
\exp\left[-(\bk-\bk_0)^2/2\Delta k^2 \right],
\end{equation}
normalized to $\Tr \rho_0= \int \rho_0(\bk) \rmd\bk/(2\pi)^d =1$. From time $t=0$ onwards, the matter wave then evolves according to the Schr\"odinger equation with single-particle Hamiltonian $H = \bp^2/2m+V(\br)$. A well-controlled random potential is provided by laser speckle \cite{Clement2006}. Without loss of generality, $\av{V(\br)}=0$, where the overbar denotes the ensemble average over disorder realizations. The random potential is then characterized by its variance $\av{V(\br)^2}=V^2$ and spatial correlation length $\zeta$. This length defines a correlation time $\tau_{\zeta}= m\zeta^{2}/\hbar$ and a correlation energy $E_{\zeta}=\hbar^{2}/(m\zeta^{2})$ \cite{Kuhn}.

First, we study the dynamics of the matter wave expanding in a repulsive 2D speckle potential by solving numerically the Schr\"{o}dinger equation for a potential strength $V=E_{\zeta}$, and initial condition $k_{0}\zeta=2$, $\Delta k=0.01 k_0/\sqrt{2}$. For $^{87}$Rb atoms in a correlated potential with $\zeta = 1\,\mu$m, the absolute time scale is $\tau_{\zeta}= 1.365\,$ms. Fig.~\ref{cbs} shows the numerically computed, ensemble-averaged momentum distribution $\overline{\rho}(k_x,k_y,t)$ at two different times.

At short times [Fig.~\ref{cbs}(a), $t=10\tau_\zeta$], one sees a very narrow peak at $\bk_0$, a broad, ring-shaped anisotropic background and a rather smooth peak at $-\bk_0$. The forward peak is the remainder of the initial momentum distribution, which is depleted because atoms are scattered out of the initial mode at a rate given by the elastic scattering time $\tau_s$ \cite{Montambaux}. The latter can be extracted from the early-time decay $\overline{\rho}(\bk_0,t)\approx\left|\overline{\Psi}(\bk_0,t)\right|^2\propto\exp(-t/\tau_s)$. We find $\tau_s = 1.43\tau_\zeta$, with corresponding mean free path $\ell_s=v_0\tau_s$ such that $k_0\ell_s =  5.72$ for the parameters used. Weak-disorder perturbation theory \cite{Kuhn} predicts too low a value ($k_0\ell_s=2.32$), as known for rather strong, spatially correlated disorder \cite{Hartung08}. This shows how the early-time momentum distribution can be used to measure the key parameter $\tau_s$, even in the strong disorder regime where precise analytical predictions are not available \cite{Footnote}.

Atoms scattered out of the initial mode populate all other accessible $k$-space modes on the energy shell and thus appear along the circle $|\bk| = |\bk_0|$ in Fig.\ \ref{cbs}. Due to disorder broadening, the energy shell has a finite width, of order $\ell_s^{-1}$, which is larger than the initial width $\Delta k$ for the chosen parameters. After a time of the order of the Boltzmann time $\tauB =8.5\tau_\zeta$ ($k_0\lB = 34$) \cite{Kuhn}, the dynamics turns from ballistic to diffusive. As the memory about the initial direction of propagation gets erased, the diffusive momentum distribution then becomes isotropic on average. More precisely, for $t\gtrsim\tau_B$, i.e. when diffusion is fully established, we find that the decay of the anisotropic Fourier components of the background is well fitted by $\exp(-t/\tau)$, where $\tau$ is the transport time that governs the diffusive dynamics. For the present parameters, one finds $\tau =  5.8\tau_\zeta$ (corresponding to a transport mean free path $k_0\ell = 23.2$). As expected, $\tau$ is smaller than $\tauB$, due to weak localization (WL) corrections \cite{Hartung08} arising at early times and caused by very short CBS loops. In Fig.\ \ref{cbs}(a), the peak at $-\bk_0$  is the incipient CBS signal. 

\begin{figure}

{\includegraphics[scale=0.73]{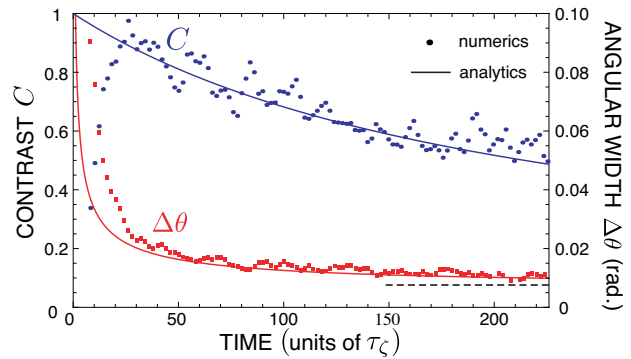}}
\caption{(Color online) CBS peak contrast $C$ (blue circles) and angular width $\Delta\theta$ (red squares) as obtained from the numerics after averaging over $960$ disorder configurations and over a time window of $10\tau_{\zeta}$. Solid curves: theoretical predictions \eqref{C_t} and \eqref{Dtheta_t}, respectively. Dashed horizontal line: angular width $\Delta\theta_0=0.01/\sqrt{2}$ of the initial momentum distribution, asymptotically reached by the CBS width for times much larger than the coherence time $\tau_\Delta = (2D\Delta k^2)^{-1} = 216\tau_\zeta$.} \label{figcontrast}
\end{figure}

At longer times [Fig.~\ref{cbs}(b), $t=18\tau_\zeta$], the initial state is totally depleted, the diffusive background is fully isotropic, and the CBS peak is the dominant feature. Both its contrast $C$, defined as the height above the diffusive background, and its angular width $\Delta\theta$ slowly decrease with time, as shown in Fig.~\ref{figcontrast}.


\begin{figure}[b]
{\centering{\includegraphics[scale=0.41]{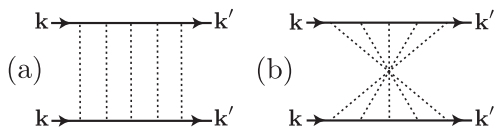}}}
\caption{Ladder (a) and maximally crossed (b) 
multiple scattering Feynman diagrams, giving rise to the diffusive and CBS contributions, Eq.\ \eqref{Phi_diffusion}  and Eq.\ \eqref{PhiC}, respectively. \\
}
\label{diagrams}
\end{figure}

For a quantitative understanding of these observations, we now turn to the analytical description of matter wave dynamics in dimension $d=2,3$. The ensemble-averaged momentum distribution $\av{\rho}(\bk',t)$ at time $t$ is given by \cite{Kuhn}
\begin{equation} \label{rhokprimet}
\av{\rho}(\bk',t) =  \intdd{\bk} \intdone{E}\Phi_{\bk\bk'E}(0,t)\rho_0(\bk), 
\end{equation}
where the intensity propagation kernel $\Phi_{\bk\bk'E}$ 
projects the initial momentum $\bk$ on the energy shell $E$, describes the ensuing unitary dynamics generated by the Hamiltonian $H$, and projects back onto the final momentum $\bk'$. For long enough times $t\gg\tauB$, but well before the onset of Anderson localization, the atomic dynamics is diffusive with an energy-dependent diffusion constant $D(E)=2E\tau/(md)$ that incorporates the short-range WL corrections. The intensity propagation kernel then takes the form 
\begin{equation}\label{Phi_diffusion}  
\Phi_{\bk\bk'E}^{(\text{L})}(\bq,t) = 
\frac{A(\bk,E)A(\bk',E)}{2\pi \nu(E)}\exp[-D(E)\bq^2 t ].
\end{equation} 
The spectral function $A(\bk,E)= 2\pi \bra{\bk}\av{\delta(E-H)}\ket{\bk}$ is the average probability density that a plane-wave state $\ket{\bk}$ has energy $E$. It also determines the average density of states $\nu(E) = \int A(\bk,E)\rmd\bk/(2\pi)^{d+1}$. Using \eqref{Phi_diffusion} at momentum transfer $\bq=0$ in Eq.\ \eqref{rhokprimet} results in a time-independent isotropic diffusive background: 
\begin{equation} \label{rhokprimeL}
\av{\rho}^{(\text{L})}(\bk') =  \intdd{\bk} \intdone{E} \frac{A(\bk,E)A(\bk',E)}{2\pi \nu(E)}\rho_0(\bk).
\end{equation} 
This contribution to the momentum distribution describes scattering processes that do not rely on long-range phase coherence; the label ``L'' refers to the ladder topology of the underlying Feynman diagrams, Fig.\ \ref{diagrams}(a) \cite{Montambaux}.
Phase-coherent multiple scattering leads to an additional contribution, given by the maximally crossed diagrams shown in Fig. \ref{diagrams}(b). These diagrams describe the interference of amplitudes that propagate along the same set of scatterers, but in opposite directions, and give rise to the CBS effect. Its contribution $\av{\rho}^{(\text{C})}(\bk',t)$ is obtained from Eq.\ \eqref{rhokprimet} after the substitution $\bq=\bk+\bk'$ in Eq.\ \eqref{Phi_diffusion}, 
\begin{equation}  
\label{PhiC}
\Phi^{(\text{C})}_{\bk\bk'E}(0,t) = 
\Phi^{(\text{L})}_{\bk\bk'E}(\bk+\bk',t). 
\end{equation} 
Because this coherent kernel peaks at $\bk'=-\bk$, the CBS signal will appear at $-\bk_0$ if the initial distribution is centered around $\bk_0$, as borne out beautifully by Fig.~\ref{cbs}. For an initial plane-wave state ($\Delta k=0$), the CBS contribution at exact backscattering is at all times exactly equal to the background level, and thus enhances the diffusive density at that point by a factor of two. This has been observed in optical and acoustical experiments \cite{Wolf85, Fink97}.

If, however, the initial state has a finite momentum spread, the CBS signal is reduced because Eq.~\eqref{rhokprimet} convolves the plane-wave kernel \eqref{PhiC} with the initial momentum distribution. In order to be able to detect other dephasing processes, we therefore need to analyze in detail the CBS signal for matter wave packets with finite momentum spread. From here on, we assume 
\begin{equation} \label{regime_of_interest}
k_0^{-1}\ll \ell_s \le \ell \ll \Delta k^{-1}.
\end{equation} 
The first inequality implies the weak-disorder condition  $k_0\ell_s \gg1$ \cite{Montambaux}, whereas the last one requires the atom coherence length $\Delta k^{-1}$ to span many mean free paths, a necessary condition to observe the CBS peak (see below).

Let us first discuss the diffusive contribution \eqref{rhokprimeL}. The regime of interest (\ref{regime_of_interest}) implies $\Delta k\ll \ell_s^{-1}$, i.e.\ an initial distribution $\rho_0(\bk)$  much narrower than the spectral density. This is apparent in Fig.\ \ref{cbs}(a) where the initial peak is much narrower than the diffusive background. Therefore, the result of their convolution in \eqref{rhokprimeL} can be approximated by $A(\bk_0,E) \approx A_\gamma(E_0-E)=\hbar\gamma/[(E_0-E)^2+\hbar^2\gamma^2/4]$, where $\gamma=\tau_s^{-1}$ and $E_0=E_{k_0}= \hbar^2k_0^2/(2m)$ \cite{Kuhn, Montambaux}. The remaining integration over $E$ gives 
\begin{equation}\label{Background_final}
\av{\rho}^{(\text{L})}(\bk') =
\frac{A_{2\gamma}(E_{k'}-E_0)}{2\pi\nu(E_0)}. 
\end{equation} 
The background reaches its maximum $\av{\rho}^{(\text{L})}(\bk_0) =\tau_s/[\pi \hbar\nu(E_0)]$ on shell. Spectral broadening results in a relative half-maximum width of $\hbar\gamma/E_0 =2/(k_0\ell_s)$.

Consider now the CBS contribution, Eq.\ \eqref{rhokprimet} with \eqref{PhiC}. Since $D(E)$ is a smooth function of $E$ on the scale of $\hbar\gamma$, one can approximate 
\begin{equation}
\label{PhikkpCBSap}
\intdone{E}\Phi^\text{(C)}_{\bk\bk'E}(t) = \frac{A_{2\gamma}(E_k-E_{k'})}{2\pi \nu(E_k)} e^{-D(E_k)(\bk+\bk')^2 t}.
\end{equation} 
The exponent on the right-hand side introduces the CBS $k$-space width $\Delta k_c = 1/\sqrt{2Dt}$, which is the inverse of the diffusive spread in real space. Written as $\Delta k_c = \ell^{-1} \sqrt{\tau d/(2t)}$, this width is seen to be \emph{always smaller} than the spectral width $\ell_s^{-1}$ in the diffusive regime $t\gg\tauB$. It is therefore \emph{never permissible} to approximate the spectral function $A_{2\gamma}(E_k-E_{k'})$ by its clean limit $2\pi\delta(E_k-E_{k'})$ when calculating the CBS peak resulting from a wave packet of finite width.

Within the regime delimited by inequalities (\ref{regime_of_interest}), the CBS contribution at $\bk'=-\bk_0+\bq$ around backscattering then follows directly by Gaussian integration,  
\begin{equation}
\label{CBS_final}
\av{\rho}^\text{(C)}(\bq-\bk_0,t) =  \av{\rho}^{(\text{L})}(\bk_0)\times 
\frac{ \exp[-\bq^2/2\Delta q^2(t) ] }{(1+t/\tau_\Delta)^{d/2}} .
\end{equation}
Here, $\tau_\Delta=(2 D \Delta k^2)^{-1}$ is the time it takes the matter wave to spread diffusively over the coherence length. The CBS contrast $C(t)=\av{\rho}^\text{(C)}(-\bk_0,t)/\av{\rho}^{(\text{L})} (\bk_0)$ is found to decrease like
\begin{equation}
\label{C_t}
C(t)= (1+t/\tau_\Delta)^{-d/2}. 
\end{equation}
The CBS signal can only be observed if this contrast is not too small at time $\tau$. The corresponding condition $\tau_\Delta\gg\tau$ requires the initial wave packet to be coherent over a distance $\Delta k^{-1} \gg \ell$, as stated in \eqref{regime_of_interest}. As function of $\bq$, the diffusive CBS signal \eqref{CBS_final} is an isotropic Gaussian. Its variance $\Delta q^2(t)=\Delta k^2 + (2D t)^{-1} = \Delta k^2 (1+\tau_\Delta/t$) is the sum of the variances of the initial momentum distribution \eqref{rho0iso} and the interference kernel \eqref{PhikkpCBSap}. Expressed as a function of the angle $\theta$ away from backscattering at fixed radius $k'=k_0$, the CBS peak width decreases as 
\begin{equation}
\label{Dtheta_t}
\Delta\theta(t)=\Delta\theta_0 \sqrt{1+\tau_\Delta/t},
\end{equation}
where $\Delta\theta_0 = \Delta k/k_0$ is the angular width of the initial state in momentum space. In Fig.\ \ref{figcontrast}, the predictions \eqref{C_t} and \eqref{Dtheta_t} for $d=2$ are plotted together with the numerical data ($\tau_\Delta =  216\tau_\zeta$). The agreement is excellent in the diffusive regime, validating our analytical description of matter wave CBS in bulk disorder.


We now discuss the behavior of the CBS peak at the Anderson
localization onset. In 2D, this occurs when the diffusive
spread $\sqrt{2Dt}$ reaches the localization length $\xi = \lB
\exp(\pi k_0\lB/2)$, i.e.\ around the time $\tau_\text{loc}
=\xi^2/D$. This regime is not visible in Fig.~\ref{figcontrast} since
$\tau_\text{loc}=10^{24}\tau_\zeta$ for the parameters used. 
Applying the self-consistent theory of localization
  \cite{Vollhardt92}, we predict two possible scenarios. If the
initial coherence length is much smaller than the localization length,
$(\Delta k)^{-1} \ll \xi$, the CBS peak width saturates  at
$\Delta\theta_0$ already around the time $\tau_\Delta
\ll\tau_\text{loc}$. The contrast continues to decrease
like $(\tau_\Delta/t)^{d/2}$ until it reaches $C_\text{loc} = (
\tau_\Delta/\tau_\text{loc})^{d/2} \ll1$. If, on the contrary, the
coherence length exceeds the localization length, $(\Delta k)^{-1} \gg
\xi$, then the CBS interference becomes sensitive to the localization
of the wave amplitudes around the time
$\tau_\text{loc}\ll\tau_\Delta$, and the width freezes at
  $\Delta\theta_\text{loc}=(k_0\xi)^{-1}$, with contrast, Eq.~\eqref{C_t},
  of order unity.
Observing this effect in an
actual experiment would provide compelling evidence for Anderson
localization, and the CBS peak could serve as an independent measure
for the localization length. Supplementary evidence would be gained by
perturbing the phase coherence deliberately, for example by varying
the speckle potential over the course of time, in order to dephase the
CBS signal and suppress localization in a controlled manner.  

Let us conclude by showing that the proposed experiment is
  immediately feasable with current techniques. 
  Ref.\cite{Jendrzejewski11} uses condensed bosons at
  finite temperature, with a momentum spread $(\Delta k)^{-1}$ of thermal and condensed
  components that we estimate at $2.5\,\mu$m and $1.6\,\mu$m,
  respectively. To ensure a good CBS contrast, these values should not be
  smaller than $\ell$, according to Eq.~\eqref{regime_of_interest}.  
For our choice of $k_0=2/\zeta$, we
  find $\ell\simeq2.3\,\mu$m and $k_0^{-1}\simeq0.1\,\mu$m  for 
  $\zeta\simeq0.2\,\mu$m of Ref. \cite{Jendrzejewski11}, such that
  $k_0^{-1}\ll\ell\sim (\Delta k)^{-1}$. We thus predict that for a
  slightly more coherent cloud or a slightly smaller $k_0$
  under the same conditions, CBS should be
  clearly observable. Finally, we briefly comment on atomic
  interactions, leaving their detailed study for future
  investigation. Their primary effect is to decrease the CBS contrast
  \cite{Hartung08}. For a ratio of order $10$ between disorder
  strength and atomic interaction energy, as in
  Ref.~\cite{Jendrzejewski11}, preliminary numerical calculations show
  a reduction of $C$ by only $10\%$ at time $t=18\tau_\zeta$. The CBS
  effect of matter waves is therefore a robust phenomenon that offers
  a unique opportunity to characterize key transport parameters and to
  monitor the transition from diffusion to Anderson localization.


We thank D.~Delande, G.~Labeyrie and J.-F.~Schaff for fruitful discussions. NC acknowledges financial support from the Alexander von Humboldt Foundation and hospitality by the Centre for Quantum Technologies (CQT). ChM and BG acknowledge funding from the CNRS PICS and from the France-Singapore Merlion programs. The CQT is a Research Centre of Excellence funded by the Ministry of Education and the National Research Foundation of Singapore.


\begin{thebibliography}{99}
\bibitem{Footnote}
Corrections to the lowest-order result for $\tau_s$ in a speckle potential involve non-Gaussian terms \cite{Kuhn} that are not captured by the self-consistent Born approximation \cite{Skipetrov08}.
\end{thebibliography}
\end{document}